\documentstyle[twocolumn,aps]{revtex}

\begin{document}
\draft
\title{Tensor form of magnetization damping }
\author{Vladimir L. Safonov}
\address{Center for Magnetic Recording Research, \\
University of California - San Diego, \\
9500 Gilman Drive, La Jolla, CA 92093-0401}
\date{\today}
\maketitle

\begin{abstract}
A tensor form of phenomenological damping is derived for small magnetization
motions. This form reflects basic physical relaxation processes for a
general uniformly magnetized particle or film. Scalar Landau-Lifshitz
damping is found to occur only for two special cases of system symmetry.
\end{abstract}


\section{Introduction}

The dynamics of magnetization ${\bf M}$ of a single-domain ferromagnetic
particle is usually described by the Landau-Lifshitz (LL) equation \cite
{landau}:

\begin{equation}
\frac{d{\bf M}}{dt}=-\gamma {\bf M}\times {\bf H}_{{\rm eff}}-\frac{\alpha
\gamma }{M_{{\rm s}}}{\bf M}\times ({\bf M}\times {\bf H}_{{\rm eff}}).
\label{principal}
\end{equation}
Here ${\bf H}_{{\rm eff}}$ is the effective field, $\gamma $ is the
gyromagnetic ratio, $M_{{\rm s}}=|M|$ is the saturation magnetization and $%
\alpha $\ is a dimensionless damping parameter. The first term in (\ref
{principal}) directly follows from microscopic equations and describes an
averaged precession of a large number of spins. The second term in (\ref
{principal}) was introduced phenomenologically in Ref.\cite{landau} just
from a\ simple geometric consideration to describe the magnetization
damping. Later Gilbert \cite{gilbert} declared that this damping term can be
rewritten as a ``dry friction''. Here we focus on the case of small damping (%
$\alpha \ll 1$) and neglect non-uniform magnetization motions.

More complicated forms of phenomenological damping were proposed by
Bar'yakhtar and co-authors \cite{baryakhtar},\cite{baryakhtar1} on the base
of general symmetry considerations of exchange and relativistic relaxation
processes. According to these papers, for a uniform magnetization motion
``the crystal symmetry should influence the form of the relaxation terms''
and, therefore the phenomenological damping term should contain a damping
tensor with several damping parameters (``hierarchy of dissipative terms'')
instead of one (isotropic) damping.

However, a general problem of magnetization damping can not be solved using
just symmetry considerations. The magnetization relaxation process appears
as a result of microscopic interactions of spins with each other and with
phonons, conduction electrons and so on. In other words, a direct connection
with microscopic physics must be found for the damping terms. The
microscopic derivation of damping has been performed, for example, in the
case of ``valence-exchange'' relaxation \cite{kambersky} and magnetization
relaxation on paramagnetic impurities \cite{safbert3}. A tensor structure of
damping term follows from the results of Refs \cite{kambersky} and \cite
{safbert3}.

The aim of this paper is to demonstrate that the conventional relaxation
term in the LL equation is inconsistent (with the exception of two special
cases)\ with a form that follows from basic equations and derive a new form
of phenomenological damping. This new tensor form of damping contains only
one phenomenological parameter, which can be found experimentally. The
dynamic equation with tensor damping reflects physical relaxation processes
for a general uniformly magnetized particle or film. The length of the
magnetization vector is conserved and a rate of change of the energy
proportional to the square of the torque.

\section{Small magnetization motion}

Let us consider a uniformly magnetized ferromagnetic particle. We shall
study small-amplitude motions of magnetization in the vicinity of
equilibrium state ${\bf M=}M_{{\rm s}}\widehat{{\bf z}}_{1}$, where $%
\widehat{{\bf z}}_{1}$ is the unit vector in the equilibrium direction. It
is well-known that if we neglect the loss of energy, the magnetization
rotation around effective field in the vicinity of equilibrium, in general,
is elliptical. If $\widehat{{\bf x}}_{1}$ and $\widehat{{\bf y}}_{1}$ are
the unit vectors corresponding to principal ellipse axes, the magnetic
energy ${\cal E}$ can be written in the form:

\begin{equation}
{\cal E}/V=\frac{H_{1}}{2M_{{\rm s}}}M_{x_{1}}^{2}+\frac{H_{2}}{2M_{{\rm s}}}%
M_{y_{1}}^{2}.  \label{energy}
\end{equation}
Here $V$ is the particle volume, $H_{1}$ and $H_{2}$\ are positive fields,
which include both microscopic and shape anisotropies and the external
magnetic field.

Our aim now is to derive equations of motion for $M_{x_{1}}$ and $M_{y_{1}}$%
. We can do this by two different methods: a) with the help of
Landau-Lifshitz Eq.(\ref{principal}) and b) with the help of a normal mode
approach, where the relaxation is introduced from basic equations.

\subsection{Linearized LL equations}

Using (\ref{energy}) and (\ref{principal}), we can calculate the effective
field ${\bf H}_{{\rm eff}}=-\partial ({\cal E}/V)/\partial {\bf M}$ and
write down the linearized equations for the transverse magnetization
components ($M_{z_{1}}\simeq M_{{\rm s}}$):

\begin{equation}
\frac{d}{dt}\left( 
\begin{array}{c}
M_{x_{1}} \\ 
M_{y_{1}}
\end{array}
\right) =\left( 
\begin{array}{cc}
-\alpha \gamma H_{1} & -\gamma H_{2} \\ 
\gamma H_{1} & -\alpha \gamma H_{2}
\end{array}
\right) \left( 
\begin{array}{c}
M_{x_{1}} \\ 
M_{y_{1}}
\end{array}
\right) .  \label{mxmyLLG}
\end{equation}
In the absence of relaxation ($\alpha =0$), the diagonal terms in (\ref
{mxmyLLG}) are equal to zero and the frequency of ferromagnetic resonance is:

\begin{equation}
\omega _{0}=\gamma \sqrt{H_{1}H_{2}}.  \label{FMR}
\end{equation}

According to Eq. (\ref{mxmyLLG}), the diagonal terms, responsible for
relaxation, in general, are different ($H_{1}\neq H_{2}$). The damping
coefficients are equal only in two special cases when $H_{1}=H_{2}$: 1) the
case of spherical symmetry and 2) the case of uniaxial symmetry when the
external magnetic field and equilibrium magnetization are oriented along the
easy axis.

\subsection{Normal mode approach}

In this approach it is convenient to introduce the classical spin ${\bf S}=-%
{\bf M}V{\bf /}\hbar \gamma $. Thus the energy (\ref{energy}) becomes

\begin{equation}
{\cal E}/\hbar =\frac{\gamma H_{1}}{2S}S_{x_{1}}^{2}+\frac{\gamma H_{2}}{2S}%
S_{y_{1}}^{2}.  \label{energy1}
\end{equation}
We shall describe small oscillations of the magnetization in terms of
complex variables$\;a^{\ast },\,a$ which are classical analogs of creation
and annihilation operators of a harmonic oscillator and can be introduced by
a Holstein-Primakoff transformation \cite{hopri} for $S_{z}\approx -S$: 
\begin{equation}
S_{x_{1}}\simeq (a^{\ast }+a)\frac{\sqrt{2S}}{2},\quad S_{y_{1}}\simeq
(a^{\ast }-a)\sqrt{2S}/2i.  \label{hp-a}
\end{equation}
The energy ({\ref{energy1}}) now can be rewritten in the form

\begin{equation}
{\cal E}/\hbar ={\cal A}a^{\ast }a+({\cal B}/2)(aa+a^{\ast }a^{\ast }),
\end{equation}
where ${\cal A}=\gamma (H_{1}+H_{2})/2$ and ${\cal B}=\gamma (H_{1}-H_{2})/2$%
.

The dynamic precession equations are given by

\begin{equation}
da/dt=-i{\cal A}a-i{\cal B}a^{\ast },\quad da^{\ast }/dt=i{\cal A}a^{\ast }+i%
{\cal B}a.  \label{eq-a}
\end{equation}
The mixed terms in (\ref{eq-a}) can be eliminated by the linear canonical
transformation 
\begin{eqnarray}
a=uc+vc^{\ast }, &\quad &a^{\ast }=uc^{\ast }+vc,  \label{canonic} \\
u=\sqrt{\frac{{\cal A}+\omega _{0}}{2\omega _{0}}}, &\quad &v=-\frac{{\cal B}%
}{|{\cal B}|}\sqrt{\frac{{\cal A}-\omega _{0}}{2\omega _{0}}}.  \nonumber
\end{eqnarray}
Thus we describe the precession in terms of the normal mode ($c$, $c^{\ast }$%
) with energy of an harmonic oscillator 
\begin{equation}
{\cal E}/\hbar =\omega _{0}c^{\ast }c,  \label{ener}
\end{equation}
where $\omega _{0}=\sqrt{{\cal A}^{2}-{\cal B}^{2}}$ is the frequency of
ferromagnetic resonance equivalent to Eq.(\ref{FMR}) (${\cal A}+{\cal B}%
=\gamma H_{1}$ and ${\cal A}-{\cal B}=\gamma H_{2}$). The dynamic equations
for $c$ and $c^{\ast }$ are now independent:

\begin{equation}
dc/dt=-i\omega _{0}c,\quad dc^{\ast }/dt=i\omega _{0}c^{\ast }.
\label{c-dyn}
\end{equation}

In order to construct a damped motion for this oscillator it is necessary to
consider the interaction with a thermal bath. This implies an introduction
of microscopic interactions with magnons, phonons, etc. and, in general,
this problem must be solved using quantum statistical methods. Here, for
simplicity, we will not focus on a particular microscopic relaxation
mechanism. Such a problem in a general form was solved by Lax \cite{lax}. We
shall use this result with some brief explanation.

Let us consider the Hamiltonian of harmonic oscillator interacting with a
thermal bath:

\begin{equation}
{\cal H}=\hbar \omega _{0}b^{\dagger }b+i\hbar (b^{\dagger }g-bg^{\dagger })+%
{\cal H}_{TB}.  \label{laxH}
\end{equation}
Here $b^{\dagger }$ and $b$ are the creation and annihilation bose
operators, ${\cal H}_{TB}$ is the thermal bath Hamiltonian, $g$ and its
hermitian conjugate $g^{\dagger }$ denote the thermal bath operators, which
describe a weak interaction with the oscillator. Lax analyzed the density
matrix equation with the Hamiltonian (\ref{laxH}) and obtained dynamic
equations for classical amplitudes $\langle b\rangle $ and $\langle
b^{\dagger }\rangle $ to second order in the interaction, where $\langle
...\rangle $ is the thermal bath averaging. Denoting $c=\langle b\rangle $
and $c^{\ast }=\langle b^{\dagger }\rangle $, we can write these equations as

\begin{eqnarray}
(d/dt+\eta )c &=&-i(\omega _{0}+\Delta \omega )c,  \label{normalmode} \\
(d/dt+\eta )c^{\ast } &=&i(\omega _{0}+\Delta \omega )c^{\ast },  \nonumber
\end{eqnarray}
where

\begin{eqnarray}
\eta -i\Delta \omega &\equiv &\int_{0}^{\infty }du\ e^{-i\omega
_{0}u}\langle \lbrack g(0),g^{\dagger }(u)]\rangle ,  \label{relax} \\
g(u) &=&\exp (iu{\cal H}_{TB}/\hbar )g\exp (-iu{\cal H}_{TB}/\hbar ). 
\nonumber
\end{eqnarray}
$[...,...]$ is the commutator, $\eta $ is the relaxation rate and $\Delta
\omega $ is the frequency shift due to interaction with the thermal bath
(usually $|\Delta \omega |\ll \omega _{0}$). The most important fact is that
the equations (\ref{normalmode}) for classical complex amplitudes of damped
harmonic oscillator are general. We can use these equations even if we do
not know a microscopic relaxation mechanism and find $\eta $ and $\Delta
\omega $ from experiment as phenomenological parameters.

From Eq.(\ref{normalmode}), utilizing (\ref{hp-a}) and (\ref{canonic}) with $%
{\bf M}=-\hbar \gamma {\bf S}/V$, we derive linearized equations for $%
M_{x_{1}}$ and $M_{y_{1}}$:

\begin{equation}
\frac{d}{dt}\left( 
\begin{array}{c}
M_{x_{1}} \\ 
M_{y_{1}}
\end{array}
\right) =\left( 
\begin{array}{cc}
-\eta & -\gamma H_{2} \\ 
\gamma H_{1} & -\eta
\end{array}
\right) \left( 
\begin{array}{c}
M_{x_{1}} \\ 
M_{y_{1}}
\end{array}
\right) .  \label{mxmyFirst}
\end{equation}
According to (\ref{mxmyFirst}), the damping of $M_{x_{1}}$ and $M_{y_{1}}$
are identical. Such intrinsic isotropy of the transverse damping components
is seen in the Bloch-Bloembergen relaxation term (see, e.g., \cite{bloch}).
We also see that the non-diagonal terms in (\ref{mxmyLLG}) and (\ref
{mxmyFirst}), as expected, coincide with each other. The diagonal terms,
responsible for relaxation, are different. This means that the damping term
in Landau-Lifshitz equation (\ref{principal}) is inconsistent with the basic
physics of a damped harmonic oscillator (excluding two special cases
mentioned above).

\section{Constructing the damping term}

We can generalize Eq.(\ref{principal}) to the form:

\begin{equation}
\frac{d{\bf M}}{dt}=-\gamma {\bf M}\times {\bf H}_{{\rm eff}}-\gamma \frac{%
{\bf M}}{M_{{\rm s}}}\times \lbrack \stackrel{\leftrightarrow }{\alpha }%
\cdot ({\bf M}\times {\bf H}_{{\rm eff}})].  \label{LL}
\end{equation}
Here a dimensionless damping tensor $\stackrel{\leftrightarrow }{\alpha }$
is introduced. Note that the new tensor damping conserves the length of the
magnetization vector ($|{\bf M}|=M_{{\rm s}}$) and gives a rate of change of
the energy proportional to the square of the torque ${\bf M}\times {\bf H}_{%
{\rm eff}}$.

The tensor $\stackrel{\leftrightarrow }{\alpha }$ should contain all
necessary information about symmetry of the system. Such information is
included in the expression for the energy of the system and can be expressed
as a tensor $\partial ^{2}({\cal E}/V_{0})/\partial {\bf M}\partial {\bf M}$%
. Thus, we can consider

\begin{equation}
\stackrel{\leftrightarrow }{\alpha }=\kappa \ \frac{\partial ^{2}({\cal E}%
/V_{0})}{\partial {\bf M}\partial {\bf M}}=-\kappa \ \frac{\partial {\bf H}_{%
{\rm eff}}}{\partial {\bf M}},  \label{damptensor}
\end{equation}
where $\kappa $ is a dimensionless parameter. Substituting (\ref{energy})
into (\ref{damptensor}), we obtain the damping tensor (\ref{damptensor}) in
the vicinity of equilibrium in the form:

\begin{equation}
\stackrel{\leftrightarrow }{\alpha }=\kappa \left( 
\begin{array}{ccc}
H_{1}/M_{{\rm s}} & 0 & 0 \\ 
0 & H_{2}/M_{{\rm s}} & 0 \\ 
0 & 0 & 0
\end{array}
\right) .  \label{alpha}
\end{equation}
Equation (\ref{LL}) with the tensor $\stackrel{\leftrightarrow }{\alpha }$ (%
\ref{alpha})\ must be consistent with the Eqs. (\ref{mxmyFirst}) for small
magnetization oscillations. Linearizing (\ref{LL}) and comparing with (\ref
{mxmyFirst}), we obtain

\begin{equation}
\kappa =\eta \gamma M_{{\rm s}}/\omega _{0}^{2}.  \label{kappa}
\end{equation}

In the case of two (or more) stable stationary states in the vicinity of
equilibrium we have, in general, different FMR frequencies and relaxation
rates $\omega _{01}$, $\eta _{1}$ and $\omega _{02}$, $\eta _{2}$,
correspondingly.

\section{Discussion}

In this paper a new, tensor form of damping is derived that reflects
symmetry in the magnetic system energy. The damping tensor appears from a
general form of interaction of the normal modes of the magnetic system with
a thermal bath. This leads to the identical relaxation of transverse
magnetization components, as in the Bloch-Bloembergen equations. Our
analysis is exact for small oscillations about equilibrium \cite{comment1},
but the tensor form (as an isotropic damping) may also apply to large
magnetization motions. The damping tensor is scaled by {\it only one}
phenomenological damping parameter $\eta $, which can be obtained from the
experiment. It is demonstrated that the conventional damping term in the
Landau-Lifshitz equation applies only for two cases of high symmetry. One of
the most important applications of an anisotropic damping is sure to be the
case of thermal magnetization fluctuations \cite{safbertcondmat},\cite{bjs}.
The form of damping affects, so-called, fluctuation-dissipation relation and
therefore changes the estimated noise level. More detailed study of
anisotropic magnetization damping will be published elsewhere \cite
{safonovbertram}.

Some indications that the LLG equation does not agree well with experiment
in a magnetic thin film is shown in Ref.\cite{patton}. In order to check the
validity of the above new damping form it is necessary to study
experimentally in detail the ferromagnetic resonance (frequency and
linewidth) in anisotropic magnetic systems, e.g., films. The aim is to
demonstrate that $\eta $ (instead of $\alpha $) is a primary relaxation rate
which can depend on frequency $\omega $, temperature $T$, external magnetic
field and so on.

\bigskip

It is my great pleasure to thank H. Neal Bertram for valuable discussions
and suggestions. I also would like to thank H. Suhl, C. E. Patton and V.
Kambersky for helpful comments and discussions. This work was partly
supported by matching funds from the Center for Magnetic Recording Research
at the University of California - San Diego and CMRR incorporated sponsor
accounts.


\begin{references}
\bibitem{comment1}  It should be noted that the damping terms in \cite
{landau}, \cite{gilbert} and \cite{baryakhtar} were derived for the case of
small magnetization motions.
\end{references}
\end{document}